\documentclass[reprint,pre,...]{revtex4-2}
\usepackage{inputenc} 
\usepackage{amsmath} 
\usepackage{amssymb}
\usepackage{graphicx} 
\usepackage{caption} 
\usepackage{siunitx} 
\usepackage{breqn}
\usepackage{subcaption}
\usepackage{amsthm}
\usepackage{mathtools}
\usepackage{hyperref}
\usepackage[margin = 1in]{geometry}
\usepackage{natbib}
\newcommand{\figref}[1]{Fig. \ref{#1}}
\newcommand{\mybig}{\bBigg@{4}}

\makeatletter
\newcommand{\Biggg}{\bBigg@{3.5}}
\makeatother

\begin{document}

\begin{abstract}
The generalised Boltzmann equation which treats the combined localised and delocalised nature of transport present in certain materials is extended to accommodate time-dependent fields.     In particular, AC fields are shown to be  a means to probe the trapping and detrapping rates of materials under certain conditions. Conditions leading to dispersive transport are considered, and the signature of fractional/anomalous diffusion under AC electric fields is presented. 
\end{abstract}

\title{Analysis of a generalised Boltzmann equation for anomalous diffusion under time-dependent fields}
\author{Alex D.C. Myhill, Peter W. Stokes, Bronson Philippa, Ronald D. White}
\maketitle

\section{Introduction}
The classical transport of particles has been thoroughly explored experimentally and well explained theoretically with a probabilistic interpretation of independent particle motion leading to the diffusion and Fokker-Planck equations~\cite{Sokolov_2002}. It is characterised by a mean squared displacement scaling linearly in time. Dispersive transport is a fundamentally slower transport process with sublinear scaling of the mean squared displacement due to localisation during transport~\cite{Metzler_2001, Sokolov_2002}. This type of transport is observable in many systems with explanations ranging from charge carrier localisation in imperfections within organic semiconductors~\cite{Montroll_1975, Sibatov_2007} to electron trapping within bubbles formed in liquids~\cite{Borghesani_2002, Mauracher_2014} and obstruction in lipid transport across cell membranes~\cite{Saxton_2001}. 

Dispersive transport has been analytically modelled using multiple schemes, including drift-diffusion models~\cite{sze2012semiconductor}, fractional diffusion equations~\cite{Metzler_2001,Philippa_2011}, fractional Fokker-Planck equations~\cite{Metzler_Barkai_Klafter_1999, Barkai_2001} and random-walk formulations~\cite{Metzler_2001,Montroll_1975}. These models have given considerable insight into the nature of transport within these systems and, particularly in the case of the random-walk model, provided general results~\cite{Metzler_2001}. It was felt, however, that a model formulated in phase space would provide additional insight due to the well-developed classical theory and embed previous results within a broader framework. To this end Robson and Blumen developed a Boltzmann equation with a trapping term~\cite{robson2005analytically}. This was subsequently refined leading to Boltzmann's equation with memory of previous states and trapping~\cite{Philippa_2014,Stokes_Gen_Boltz}. This model accounted for collisions using the Bhatnagar-Gross-Krook (BGK) operator~\cite{BGK_1954}, which scatters particles to an isotropic Maxwellian velocity distribution. It introduced a convolution of a trapping distribution with the number density over time to account for localisation and the memory effect of dispersive transport~\cite{Philippa_2014}.

This generalised Boltzmann equation was analytically solved, along with its moments, in Laplace space~\cite{Stokes_Gen_Boltz}. It was shown that with particular choices of the trapping operator the fractional diffusion equation could be obtained from appropriate integrations and, when applied to organic semiconductor transport, the appropriate short and long-time behaviour was obtained, validating the approach~\cite{Philippa_2014, Stokes_Gen_Boltz}. This model was subsequently extended to include energy-dependent collision, trapping and loss rates, analysed using momentum transfer theory and used to generalise classical relations such as the Wannier and Einstein relations~\cite{Stokes_Gen_Bal, Robson_1984}.

It is suffice to say that these investigations have established that the generalised Boltzmann equation is a valuable resource in modelling transport with any non-local time effects. The question arises as to what new behaviour can be found and exploited for experimental or theoretical gain? Time-dependent fields are important in current experiments such as charge extraction by linearly increasing voltage (CELIV) and its variants~\cite{Juska_CELIV, Juska_2011}. Thus, it is worth using the generalised Boltzmann equation to model dispersive transport under arbitrary time-dependent fields. Furthermore, we shall develop our analysis bearing the idea of AC field experiments in mind, and the possibilities of extracting useful information via this method. 

It is worthwhile briefly outlining some important classical results for time-dependent AC fields as this will provide the bulk of the paper. An analysis by Robson and Makabe, using standard perturbation style techniques~\cite{Kumar_et_al_1980}, of the Boltzmann equation with the BGK operator and AC fields found interesting behaviour of the diffusion coefficient and velocity distribution~\cite{Robson_Makabe_AC}. Approximate relations and the qualitative behaviour of transport coefficients were established~\cite{ROBSON_WHITE_MAKABE}. Finally, highly accurate numerical simulations have been developed~\cite{White_Robson_Ness_2001} and used to find the transport coefficients and explore new behaviour arising from both AC electric and crossed electric and magnetic fields~\cite{White_Robson_Ness, White_2009_Recent_Advances, White_Robson_Ness_2002, Dujko2008AMT}. These investigations mainly explored the diffusion coefficient and found non-local time effects, without examining the effects of this on the current trace in an experiment.

The rest of this paper is dedicated to analysing the generalised Boltzmann equation, firstly with arbitrary time-dependent fields and then focusing on AC fields before analysing intuitively and mathematically the novel results and potential experimental implications which arise. 

Sec. \ref{sec:MOM} of the paper outlines the results of an investigation into the moments of the solution under arbitrary time-dependent potentials and discusses the physical implications of these solutions. Applications of the moments are examined in Sec. \ref{sec:APPMOM} which provide further insight into the physics and specific trapping operators are considered in Sec. \ref{sec:TRAP} which enables further analysis of the correspondence between gas phase and semiconductor phase transport. Sec. \ref{sec:NUM} employs numerical techniques and the configuration space description gained by integrating the Boltzmann equation to analyse experimental observables and implications. Finally we conclude by summarising and collating all of the knowledge gained through our investigation. 

\section{Preliminary Results} \label{sec:MOM}
\subsection{Generalised Boltzmann equation}
We use the generalised Boltzmann equation of Stokes et al.~\cite{Stokes_Gen_Boltz} with time-dependent acceleration and slightly different notation
\begin{widetext}
\begin{align}
	\left(\frac{\partial }{\partial t} + \mathbf{v} \cdot \frac{\partial }{\partial \mathbf{r}} + \mathbf{a}(t)\cdot \frac{\partial }{\partial \mathbf{v}}\right)f(t,\mathbf{r},\mathbf{v}) & = -\nu_c [f(t,\mathbf{r},\mathbf{v}) - n(t,\mathbf{r}) w(\alpha_c, v)] \nonumber \\ 
	& \mathrel{\phantom{=}} - \nu_t [f(t,\mathbf{r},\mathbf{v}) - \Phi(t) * n(t,\mathbf{r}) w(\alpha_d, v)] - \nu_l^f f(t,\mathbf{r},\mathbf{v}), \label{eq:maineqn}
\end{align} 
\end{widetext}
where $\nu_c$, $\nu_t$ and $\nu_l^f$ describe collisions, trapping and free particle loss frequencies respectively. The Maxwellian velocity distribution is given by $w(\alpha,v) = (\alpha^2/2\pi)^{3/2} \exp(-\alpha^2 v^2/2)$ and $\alpha^2 = m/k_B T$, where $k_B$ is the Boltzmann constant and $T$ is the scattering temperature. The convolution of the number density and $\Phi(t)$
\begin{equation}
	\Phi(t) * n(t,\mathbf{r}) = \int_{0}^{t} d\tau \Phi(t - \tau) n(\tau,\mathbf{r}),
\end{equation}
describes the release of particles from trapped states where $\Phi(t) = e^{-\nu_l^t t} \phi(t)$ represents the final release distribution accounting for the loss of trapped electrons at the rate $\nu_l^t$ as well as the distribution of trapping times $\phi(t)$~\cite{Stokes_2018, Stokes_Gen_Boltz}. 

\subsection{Moments of solution}
In this section we seek to extend the results of Stokes et al.~\cite{Stokes_Gen_Boltz} to time-dependent fields with the assumption of no boundary conditions. Before finding the moments we need to determine the number of particles over time. We integrate over phase space and find
\begin{align}
	\frac{dN}{dt} & = - \nu_t[N(t) - \Phi(t) * N(t)] - \nu_l^f N(t). \label{eq:numbertime}
\end{align}
This is solved in Laplace space and the solution is the same as that obtained by Stokes et al.~\cite{Stokes_Gen_Boltz} for the constant acceleration case
\begin{equation}
	N(p) = \frac{N(0)}{p + \nu_t [1-\Phi(p)] + \nu_l^f},
\end{equation}
where $p$ is the Laplace variable. Given particular trapping operators the inverse Laplace transform can be analytically or numerically found to give $N(t)$. It is important to note that a time-dependent acceleration does not change the total number of free particles. This is because the trapping and loss rates are independent of the number density of particles and, at this stage, the boundaries are infinite.
 
We now find the total phase space moments of the solution, where 
\begin{equation}
	\left< \left< \psi \right> \right> (t) = \int d^3 r \int d^3 v f(t,\mathbf{r},\mathbf{v}) \psi. \label{eq:genmoments}
\end{equation}
To evaluate \eqref{eq:genmoments} it is not necessary to define $N(t)$ explicitly; indeed, it can obscure the meaning of the expressions we find and hence it is left as $N(t)$ hereafter. The velocity moment $\left<\left<\mathbf{v}\right>\right>(t)$ is straightforward to obtain whilst higher order moments are expressed using the lower order solutions. The moments are presented below as functions of time (not in Laplace space as the representation becomes extremely messy due to $\mathbf{a}(t)$ being arbitrary)
\begin{widetext}
\begin{subequations}
\label{eqn:moments}
\begin{align}
	\left<\left<\mathbf{v} \right>\right> (t) N(t) & = \left(e^{-\nu_{\text{TOT}}t}\right) * [\mathbf{a}(t) N(t)], \\
	\left<\left<\mathbf{r} \right>\right> (t) N(t) & = \left[\frac{N(t)}{N(0)}\right] * [\left<\left<\mathbf{v}\right>\right>(t) N(t)], \\	
	\left<\left<\mathbf{vv} \right>\right> (t) N(t)  & = \left(e^{-\nu_{\text{TOT}}t}\right) * [(\mathbf{a}(t)\left<\left<\mathbf{v}\right>\right>(t) + \left<\left<\mathbf{v}\right>\right>(t)\mathbf{a}(t))N(t) + I_{2} \Psi(t) * N(t)]  \\
	& \mathrel{\phantom{=}} + \frac{N(0)}{\alpha_0^2} I_{2} e^{-\nu_{\text{TOT}}t} \nonumber, \\
	\left<\left<\mathbf{rv} \right>\right> (t) N(t) & = \left(e^{-\nu_{\text{TOT}}t}\right) * [\left<\left<\mathbf{vv} \right>\right> (t) N(t) + \left<\left<\mathbf{r}\right>\right>(t)\mathbf{a}(t)N(t)],  \\
	\left<\left<\mathbf{rr} \right>\right> (t) N(t) & = \left[\frac{N(t)}{N(0)}\right] * [(\left<\mathbf{rv} \right> (t) + \left<\left< \mathbf{vr} \right>\right> (t)) N(t)],
\end{align}
\end{subequations}
\end{widetext}
where 
\begin{subequations}
\begin{align}
	\Psi(t) & = \frac{\nu_c}{\alpha_c^2} \delta(t) + \frac{\nu_t}{\alpha_d^2} \Phi(t), \\
	\nu_{TOT} & = \nu_c + \nu_t + \nu_l^f,
\end{align}
\end{subequations}
and $\alpha_c^2 = m/k_B T_c$ and $\alpha_d^2 = m/k_B T_d$ where $T_c$ and $T_d$ are the temperatures of scattering and detrapping respectively, $\delta(t)$ is the standard Dirac delta functional and $I_2$ is the $2\times 2$ identity matrix. The moments presented in this form may not appear to hold much physical meaning but we can elucidate some information. 

We firstly note that the temperatures of scattering and detrapping only appear explicitly (and implicitly in higher order moments) in the evaluation of the velocity tensor. In particular, increasing these temperatures will increase the magnitude of the velocity tensor, reflecting the increase in the isotropic diffusion coefficient.

The second point to note is that the convolution over $e^{-\nu_{TOT} t}$ indicates scattering of the particles to the Maxwellian at a total rate of $\nu_c + \nu_t R(t)$, where 
\begin{equation}
	R(t) = \frac{\Phi(t) * N(t)}{N(t)}, \label{eq:ratiodetrap} 
\end{equation}
is the ratio of detrapping to trapping (see the next section for more details)~\cite{Stokes_Gen_Boltz}. The loss frequency does not appear here because loss acts equally on all free particles and hence does not change velocity distributions. The $\nu_t$ trapping frequency is scaled by the rate at which particles are released from traps. If, at some point in time, they are being released more rapidly than trapped then $\nu_t$ has a larger impact whilst if very few particles are being released it has a much smaller impact. 

\section{Application of Moments} \label{sec:APPMOM}
\subsection{Ratio of Detrapping to Trapping}
Stokes et al. found that the ratio of detrapping to trapping, $R(t)$, can be expressed as an implicit integral~\cite{Stokes_Gen_Boltz}. If the mean trapping time is divergent then this integral will not be evaluable and $R(t)$ has to be found from its defining equation~\eqref{eq:ratiodetrap}. It is also important to note that for $\tau \rightarrow \infty$, $R(t)$ approaches $R$ asymptotically but very slowly and one cannot assume at any stage that $R(t) = R$ for accurate calculations but this assumption can be made for intuitive understanding.

\subsection{Velocity moment}
The velocity moment was defined in \eqref{eqn:moments}. However, we can re-express it using the ratio of detrapping to trapping. This is outlined below, starting with the equation for the velocity moment before dividing through by $N(t)$, using \eqref{eq:numbertime} and \eqref{eq:ratiodetrap}:
\begin{subequations}
\begin{align}
	\left[\frac{d}{dt} + \nu_c + \nu_t +\nu_l^f \right] \left[\left< \left<\mathbf{v}\right>\right> N(t) \right] & = \mathbf{a}(t) N(t), \\
	\left[\frac{d}{dt} + \nu_c + \nu_t R(t)\right] \langle\langle \mathbf{v}\rangle\rangle(t) & = \mathbf{a}(t),
\end{align}
\end{subequations}
which enables direct solution for $\langle \mathbf{v}\rangle (t)$. If we consider the asymptotic solution when all transient terms have decayed away and $R(t) \rightarrow R$ we see
\begin{equation}
	\left<\left< \mathbf{v} \right>\right> (t) = e^{-\nu_I t} \int_{0}^{t} d\tau \mathbf{a}(\tau) e^{\nu_I \tau}, \label{eq:velmoment}
\end{equation}
where $\nu_I = \nu_c + \nu_t R$. Even whilst $R(t)$ is time-varying, this is valid as long as the exponential decays on much shorter timescales than the timescale at which $R(t)$ changes. 

In the limit of slowly varying acceleration (i.e. if the timescale, $\tau$, on which $\mathbf{a}(t)$ changes satisfies $\tau \gg 1/\nu_I$) we obtain
\begin{equation}
	\langle \langle\mathbf{v} \rangle\rangle (t) = \frac{\mathbf{a}(t)}{\nu_I}.
\end{equation}
This is a direct analogue of the classical mobility relation where the drift velocity is assumed to be proportional to the field strength. We note that the overall rate of relaxation once again includes the rate of detrapping to trapping for the reasons outlined earlier. 

\subsection{Drift velocity, temperature and diffusion tensors in AC fields}
One of the possible techniques used to probe system process timescales is through the use of an AC field with a variable frequency. We define $\mathbf{a}(t) = \mathbf{a}\cos(\omega t)$ and evaluate \eqref{eq:velmoment} to find
\begin{equation}
	\langle\langle \mathbf{v}\rangle\rangle (t) = \frac{\mathbf{a}}{\sqrt{\nu_I^2 + \omega^2}} \cos(\omega t - \psi),
\end{equation}
where
\begin{equation}
	\tan \psi = \frac{\omega}{\nu_I}.
\end{equation}
Once again we observe that the ``natural'' frequency of the system is provided by $\nu_I$. We can express this more explicitly by writing
\begin{equation}
	\langle\langle \mathbf{v}\rangle \rangle(t) = \frac{\mathbf{a}_{eff}}{\nu_I} \cos(\omega t - \psi),
\end{equation}
where
\begin{equation}
	\mathbf{a}_{eff}  = \frac{\mathbf{a}}{\sqrt{1 + \omega^2/\nu_I^2}}.
\end{equation}
This effective acceleration is the by-product of the oscillatory nature of the field. 

It is instructive to consider the limits of $\omega$. In the slowly varying limit, where $\omega \ll \nu_I$, we obtain $\langle\langle \mathbf{v}\rangle \rangle(t) = \mathbf{a}(t)/\nu_I$, i.e. the anticipated completely temporally localised solution. In the opposing limit, $\omega \gg \nu_I$, $\psi \rightarrow \pi/2$ and hence we obtain $\langle\langle \mathbf{v} \rangle\rangle (t) = (\mathbf{a}/\omega) \sin(\omega t)$. This is the completely non-local solution for which the effective acceleration vector must be integrated over time, although in this case it is a simple solution.

The temperature tensor is defined as $k_B T(t) = m\langle\langle\mathbf{vv}\rangle\rangle(t) - m\langle \langle\mathbf{v} \rangle\rangle(t) \langle\langle\mathbf{v} \rangle\rangle(t) $. Using similar methods to those applied previously, with the assumption of a slowly varying field we can show
\begin{equation}
	\frac{k_B T}{m}= \frac{I_2}{\alpha_{eff}^2} + \langle \langle \mathbf{v} \rangle \rangle(t) \langle \langle \mathbf{v} \rangle \rangle(t),
\end{equation}
where 
\begin{equation}
	\frac{1}{\alpha_{eff}^2} = \left[\frac{\nu_c}{\alpha_c^2} + \frac{\nu_t R}{\alpha_d^2}\right] \frac{1}{\nu_I}.
\end{equation}
Thus the temperature tensor is comprised of an isotropic and anisotropic contribution. The isotropic contribution comes from the ``temperatures'' of the collisions and trapping and the respective rates at which these processes occur. These are invariant of orientation and hence must contribute isotropically. The anisotropic contribution is the tensor comprised of the outer product of the drift velocity vector with itself. Once again this reduces to the result for the constant field.

Unfortunately it is not possible to find the full diffusion tensor even when the field is ``slowly'' varying without significant assumptions that are against the spirit of generality employed throughout this study. We can, however, still find the isotropic contribution exactly and show that it is
\begin{equation}
	D_{ISO} = \frac{1}{\nu_I} \frac{I_2}{\alpha_{eff}^2},
\end{equation}
which is the same as the isotropic result obtained by Stokes et al.~\cite{Stokes_Gen_Boltz}.

\section{Exponential Detrapping} \label{sec:TRAP}
At this juncture we narrow the discussion to trapping operators of particular interest. A condition of dispersive transport is that the mean trapping time is infinite~\cite{Montroll_1975,Metzler_2001}. For instance, many of the distributions examined behave as $t^{-\alpha}$ in the limit as $t\rightarrow \infty$ where $0 < \alpha < 1$. These distributions are, however, often quite complicated and it is difficult to gain insight into the nature of the solution. We can, alternatively, employ distributions with finite mean trap lengths. There are several options, for a fixed trap length a shifted Dirac delta is appropriate. The option we shall pursue here is exponential detrapping. In this case
\begin{equation}
 	\phi(t) = \nu_0 e^{\nu_0 t},
\end{equation}
where $\nu_0^{-1}$ is the mean trapping time. Ultimately the number of particles can be obtained relatively easily using Laplace transformations, as is discussed by Stokes et al.~\cite{Stokes_Gen_Boltz}.

Even with this relatively simple model the diffusion coefficient cannot be obtained exactly. Instead we find the steady state drift velocity and an approximation for the position moment over time. Ultimately this yields a highly accurate approximation for the diffusion coefficient. If we have an acceleration given by $\mathbf{a}(t) = \mathbf{a} \cos(\omega t )$ we obtain~\cite{Mathematica}
\begin{widetext}
\begin{align}
	D(t)  & = \frac{1}{\nu_I} \frac{I_2}{\alpha_{\text{eff}}^2}  
	+ \frac{(\mathbf{a}_{\text{eff}}/\sqrt{2}\nu_I)(\mathbf{a}_{\text{eff}}/\sqrt{2}\nu_I)}{\nu_I} \Biggg(1 + \frac{1 +  \nu_{\text{D}}/\nu_{\text{N}}}{2(1 + \omega^2/\nu_{\text{N}}^2)} \nonumber \\  
	& \mathrel{\phantom{=}}+ \frac{\sqrt{\frac{\omega^2}{\nu_I^2} \left(1 + \frac{\nu_I + \nu_{\text{D}}}{\nu_{\text{N}}}\right)^2 +  \frac{1}{4}\left(3 + \nu_{\text{D}}/\nu_{\text{N}} \right)^2}}{(1 + 4\omega^2/\nu_I^2)\sqrt{1 + \omega^2/\nu_{\text{N}}^2}} \cos(2\omega t - \psi - 2\xi - \eta)\Biggg), \label{eq:expdiff}
\end{align}
where 
\begin{subequations}
\begin{align}
	\nu_D & = -\nu_0 + \nu_t + \nu_l^f - \nu_l^t, \\
	\nu_N & = \sqrt{\nu_D^2 + 4 \nu_0 \nu_t},\\
	\tan \psi & = \frac{\omega}{\nu_I},\\
	\tan \xi & = \frac{2\omega}{\nu_I}, \\
	\tan \eta & = \frac{2 \frac{\omega}{\nu_{\text{N}}}\left(1 + \nu_{\text{D}}/\nu_{\text{N}}\right)+\frac{\nu_I}{\omega} \left(1 -\nu_{\text{D}}/\nu_{\text{N}}+2 \omega ^2/\nu_{\text{N}}^2\right)+\frac{\nu_I^2}{\omega\nu_{\text{N}}} \left(1 + \nu_{\text{D}}/\nu_{\text{N}} \right)}{2 \left(1 + \nu_{\text{D}}/\nu_{\text{N}}\right)+\frac{\nu_I}{\nu_{\text{N}}}  \left(1 + \nu_{\text{D}}/\nu_{\text{N}}\right)+\frac{\nu_I^2}{\omega^2} \left(\left(3 + \nu_{\text{D}}/\nu_{\text{N}}\right)+2 \omega ^2/\nu_{\text{N}}^2\right)}.
\end{align}	
\end{subequations}
\end{widetext}

The diffusion coefficient presented here is a generalisation of the result of Robson and Makabe using the BGK operator without trapping~\cite{Robson_Makabe_AC}. The first point to note is that the effective field is further scaled by a factor of $1/\sqrt{2}$. This is akin to the effective DC case as this is the root mean square effective field strength and is identical in nature to the result found previously~\cite{Robson_Makabe_AC}. The second point to note is that there is an additional anisotropic component due to the effect of trapping and detrapping.

It is interesting to note the field dependence of the diffusion coefficient. If the temperature is zero and $\omega/\nu_{\text{I}} \ll 1$ the diffusion coefficient will decrease to zero periodically. On the other hand, as the field frequency increases, the magnitude of the oscillations decreases. This is a direct result of the system not responding fast enough to the field and results in the effective DC case~\cite{Robson_Makabe_AC}. 

The phase shift also varies with the field frequency. As the field frequency increases, so does the total phase shift. In the high frequency limit, $\omega/\nu_I \gg 1$, the phase shift increases to $2\pi$. Thus, the diffusion coefficient will be in phase with $\cos(2\omega t)$ at both very low and very high frequencies, although at very high frequencies the magnitude is decreased drastically.

\section{Particle Numbers with Absorbing Boundaries} \label{sec:NUM}
\subsection{Diffusion equation}
The above analysis has provided new information about the transport coefficients. We now seek to examine the effects of the transport coefficients on the observables from an experiment, i.e. the number of free electrons (noting that this will be reflected in the conduction current). It is worth noting at this stage that actually extracting the details from an experiment may be difficult due to noise, the displacement current and other factors. Nonetheless, we seek to investigate the important effects. To do this we perform numerical simulations of the generalised diffusion equation (GDE). The GDE is obtained by integrating the generalised Boltzmann equation over all of velocity space and assuming Fick's law for the particle flux holds 
\begin{widetext}
\begin{align}
	\left[\frac{\partial}{\partial t} + \nu_{t}(1 - \Phi(t)*) + \nu_{l}^{f} \right]n(\mathbf{r},t) + \left< \left<\mathbf{v}\right>\right>(t) \cdot \frac{\partial n}{\partial \mathbf{r}} - D(t) : \frac{\partial^2 n}{\partial \mathbf{r}\partial \mathbf{r}} & = 0.
\end{align}
\end{widetext}
To reduce the complexity of the numerical evaluation whilst retaining the important aspects of the physical behaviour we reduce to one dimension and non-dimensionalise
\begin{widetext}
\begin{align}
    \left[\frac{\partial}{\partial t} + \nu_{t}^{ND}(1 - \Phi(t)*) + \nu_{l}^{f,ND} \right]n(z,t) + \frac{\left< \left<v_z\right>\right>(t)t_{tr}}{L} \frac{\partial n}{\partial z} - \frac{D(t)t_{tr}}{L^2} \frac{\partial^2 n}{\partial z^2} & = 0,
\end{align}
\end{widetext}
where we used $x'=x/L$ and $\nu^{ND} = \nu t_{tr}$ where $t_{tr}$ is a characteristic transit time for zero field frequency and is defined as $t_{tr} = 2^{1/4} \sqrt{L R}\sqrt{\nu_t^{ND}/a}$. This reduces the problem to one of changing non-dimensional collision frequencies to investigate physical effects without changing the scaled transit times etc.

\subsection{Solutions with exponential detrapping}
In this subsection we explore the solution to the generalised diffusion equation with exponential detrapping when the device is exposed to an AC field ($a(t) = a_0 \cos(\omega t)$) with perfectly absorbing boundary conditions~\footnote{The generalised diffusion equation was numerically solved as a delay differential equation. The convolution was discretized using Legendre-Gauss-Lobatto quadrature whilst a finite difference scheme was implemented using the fifth order WENO method in space~\cite{shu2009weno, SHU_1998_ENO}. Finally, the time stepping was performed using the Gauss-Legendre implicit Runge-Kutta (GLIRK) scheme with error tolerances of below 1\%~\cite{Butcher_IRK}. The transport coefficients in \eqref{eqn:moments} were found using this scheme and substituted into the GDE. }. The temperature is set to zero to explore the phenomena resulting from the oscillatory diffusion coefficient. The initial condition is an approximate Dirac delta, reflecting an injection of electrons into a device. The detrapping rate is chosen to be small relative to the transit time across the device. The frequency of the field is varied with $\omega = 0.02\pi\rightarrow 2\pi$ (non-dimensionalised).

\begin{figure} 
\centering
\includegraphics[width = \columnwidth]{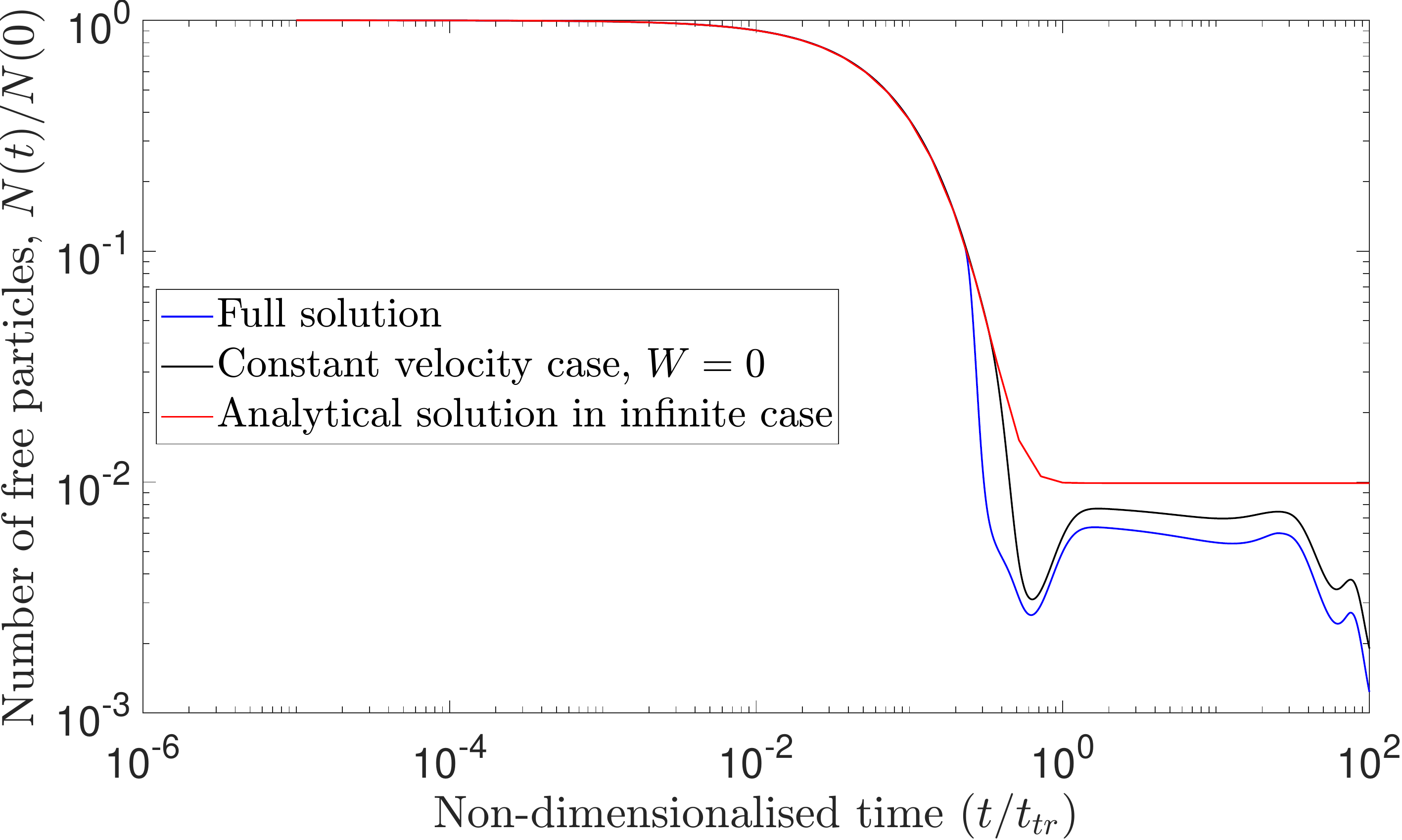}
\caption{Number of free particles remaining when an AC field with $\omega = 0.02\pi$ is applied to a semiconductor undergoing exponential detrapping given the parameters $\nu_t = 10, \nu_0 = 0.1$, $\nu_c = \nu_l^f = \nu_l^t = 0$ and $T_C = T_D = 0$ (all non-dimensionalised).}
\label{fig:exp_omega=0.1}
\end{figure}

\begin{figure} 
\centering
\includegraphics[width = \columnwidth]{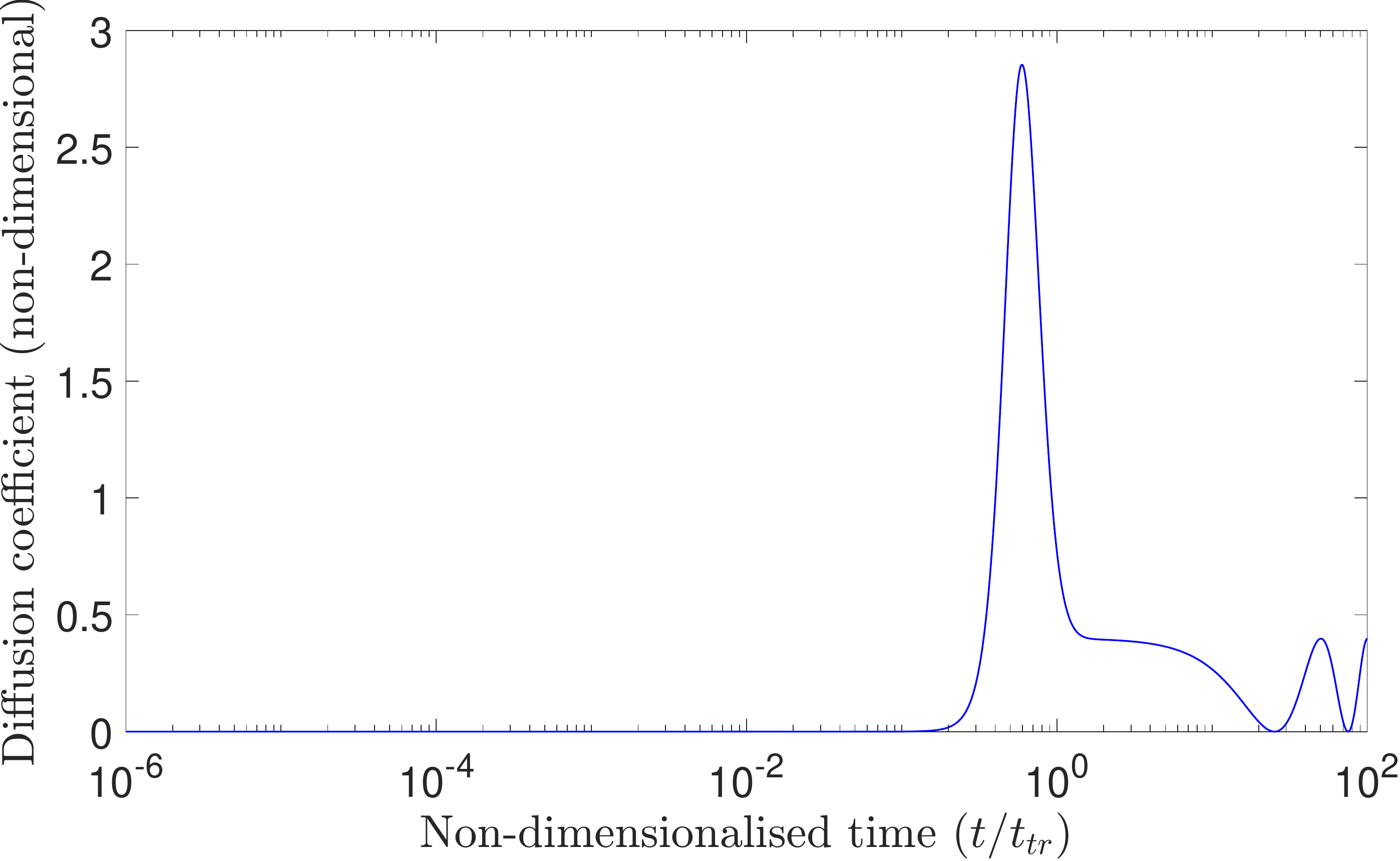}
\caption{Diffusion coefficient when an AC field with $\omega = 0.02\pi$ is applied to a semiconductor undergoing exponential detrapping given the parameters $\nu_t = 10, \nu_0 = 0.1$, $\nu_c = \nu_l^f = \nu_l^t = 0$ and $T_C = T_D = 0$ (all non-dimensionalised).}
\label{fig:diff_coeff}
\end{figure}

\begin{figure} 
\centering
\includegraphics[width = \columnwidth]{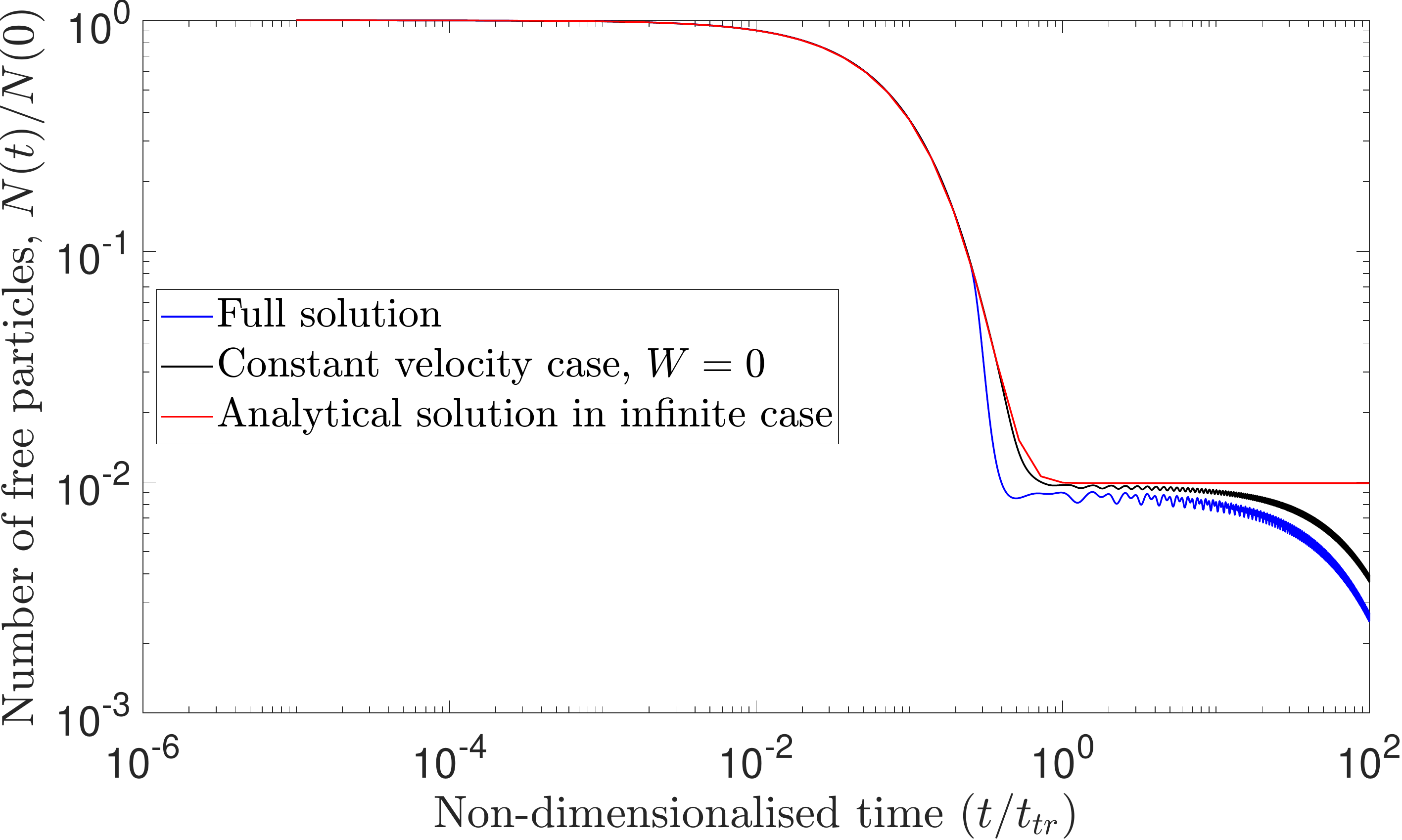}
\caption{Number of free particles remaining when an AC field with $\omega = 2\pi$ is applied to a semiconductor undergoing exponential detrapping given the parameters $\nu_t = 10, \nu_0 = 0.1$, $\nu_c = \nu_l^f = \nu_l^t = 0$ and $T_C = T_D = 0$ (all non-dimensionalised).}
\label{fig:exp_omega=10}
\end{figure}

If we compare the solutions with boundary conditions to the analytical solution without boundary conditions we see that the decrease in the number of free particles well within the first transit time is due to trapping followed by diffusion through the boundaries. At later times we observe an increase in the number of particles. Analysing the figures separately we observe that at a low applied frequency (see \figref{fig:exp_omega=0.1}) the number of free particles increases by a factor of ${\sim}3$ between $t = 0.7$ and $t = 2$. This is due to a combination of detrapping of particles (the mean detrapping time is $t = 10$) and a relatively small diffusion coefficient, which is most clearly observed when considering the diffusion coefficient shown in \figref{fig:diff_coeff}. To further underscore this point we consider the black line shown in \figref{fig:exp_omega=0.1} which displays the solution with zero drift but a time-dependent diffusion coefficient. As can be seen, similar increases in the number of free particles are observed. At higher frequencies we approach the effective DC case and the diffusion coefficient oscillations have a smaller magnitude, as is demonstrated in \eqref{eq:expdiff}. This ultimately means that the magnitude of the oscillations in the number of free particles is minimal at high frequencies, see \figref{fig:exp_omega=10}.

The final point which should be noted is observed most easily in the high-frequency regime. Here the shape of the oscillations of the number of particles in the time-dependent velocity case has a more complicated structure than in the zero drift case (where it is close to sinusoidal), which is probably due to interactions between the drift and diffusion. This would, for instance, occur with detrapping of a distribution which is non-symmetric, due to the drift of the pulse towards a boundary.

\subsection{Dispersive transport}
To model dispersive transport we use the model developed by Philippa et al.~\cite{Philippa_2014} for a Gaussian density of states
\begin{equation}
	\phi(t) = \alpha \nu_0 (t\nu_0)^{-\alpha - 1} \gamma(\alpha + 1, t\nu_0),
\end{equation}
where $\gamma(.,.)$ is the lower incomplete Gamma function and $\alpha$ is a parameter dependent on the scattering temperature and the width of the density of states. It may be expected that we will observe similar relationships between the zero field and the AC field case as observed for exponential detrapping and this is indeed the case. For both low and high field frequency cases, clearly dispersive behaviour is observed at long times with the number of particles oscillating about the trend noted by Philippa et al. of $\log N \sim -\log t$ (as opposed to the classical result of $\log N \sim -t$)~\cite{Philippa_2014}.

\begin{figure} 
\centering
\includegraphics[width = \columnwidth]{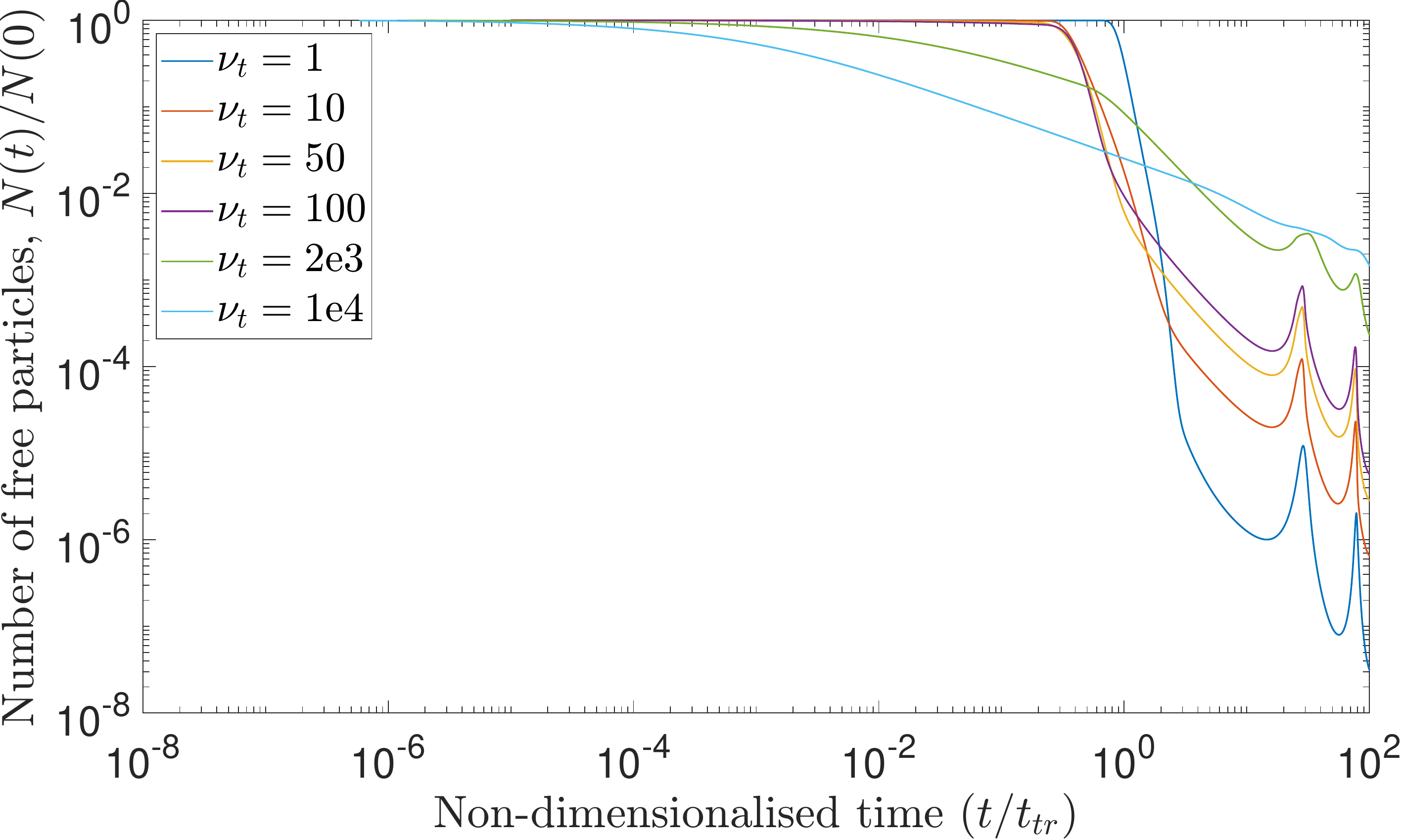}
\caption{Number of free particles remaining when an AC field with $\omega = 0.02\pi$ is applied to a semiconductor undergoing dispersive transport with the parameters $\nu_0 = 5 \times 10^{5}$, $\alpha = 0.5$, $\nu_c = \nu_l^f = \nu_l^t = 0$ and $T_C = T_D = 0$ (all non-dimensionalised).}
\label{fig:number_comparison}
\end{figure}

It was noted by Philippa et al. that the trapping frequency should be large ($\nu_t t_{tr} \gg 1$) to represent dispersive transport in organic semiconductors~\cite{Philippa_2014}. To investigate the effect of a greater trapping frequency, a simulation was run which is presented in \figref{fig:number_comparison}, comparing the effect of different trapping frequencies for $\omega = 0.02\pi$. The magnitude of the oscillations in the number of free particles reduces as the trapping frequency increases. This result is consistent with the solution approaching the zero drift, effective DC result. In this case, however, this is as a result of particles being trapped too rapidly for a large increase in free particle numbers to occur as opposed to the high field frequency. 

These results imply that an experiment at low field frequencies would give qualitative information about the trapping frequency. If the trapping frequency was high then the oscillations would be of a minimal nature, as displayed in \figref{fig:number_comparison}, whereas if the trapping frequency was quite low then the oscillations may be of a large magnitude. This may allow the trapping rate to be estimated in an experimental setting. Clearly, however, obtaining these results is going to be difficult due to the factors mentioned above. The most important of these are the magnitude of the displacement current and the temporal resolution needed to measure these effects. 

\subsection{Beer-Lambert law initial condition}
We have analysed the numerical solutions to the generalised diffusion equation with a Dirac delta initial condition and gained insight into the behaviour. We now more closely replicate a classical time-of-flight experiment with the DC field replaced by an AC field and the appropriate waiting time distribution. The experimental setup is a device placed between two perfectly absorbing electrodes. A laser pulse through one of the electrodes generates electrons and holes. The simplest model for the number density is the Beer-Lambert law (BLL)~\cite{Juska_2011}. 

Using a smoothed BLL as the initial condition with perfectly absorbing boundaries we performed numerical simulations on both the electron and hole number density. Note that the assumption of non-interaction is consistent with our assumption that number densities are low enough that space-charge effects are ignorable. 

\begin{figure} 
\centering
\includegraphics[width = \columnwidth]{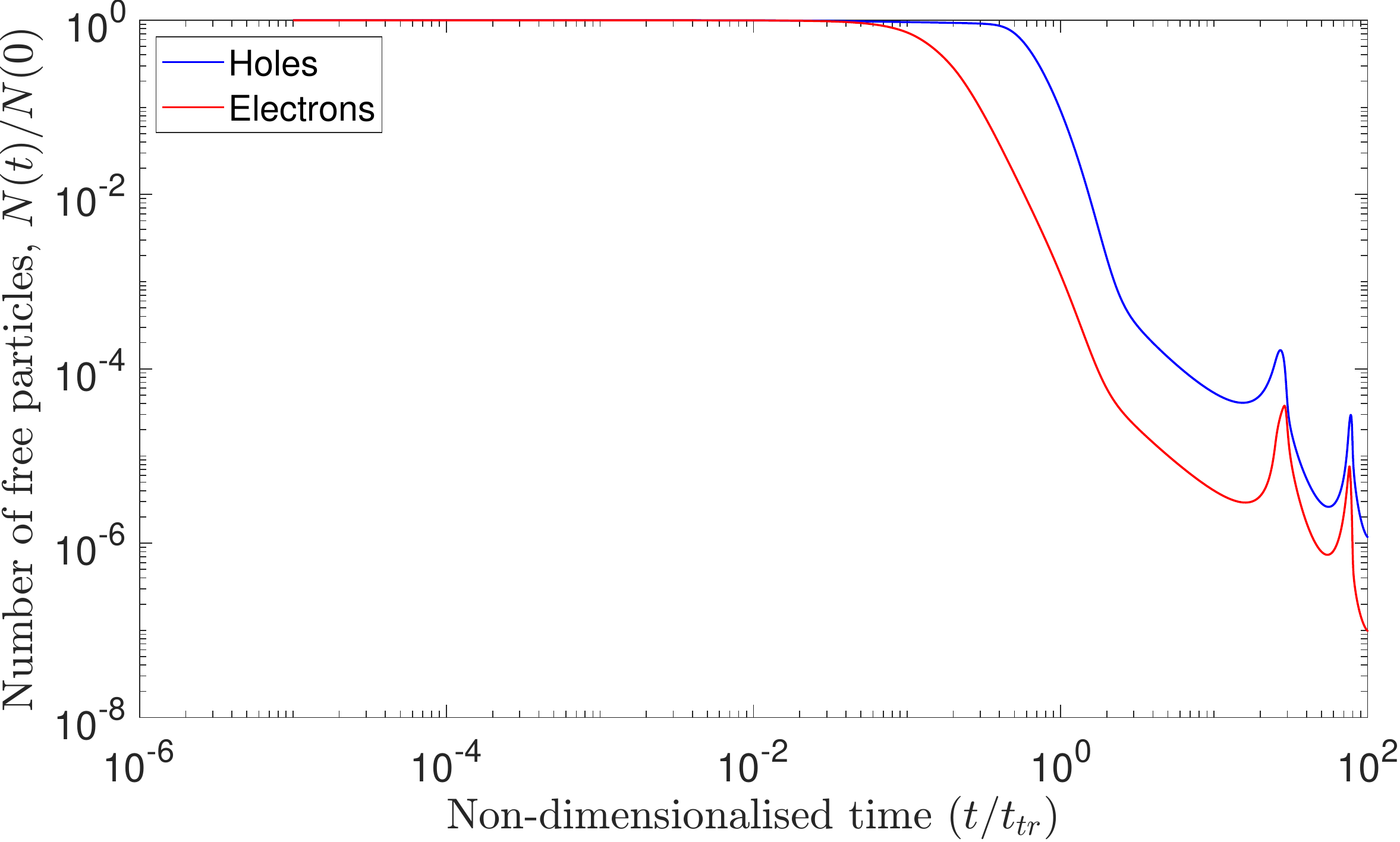}
\caption{Number of free particles remaining when an AC field with $\omega = 0.02\pi$, is applied to a semiconductor undergoing dispersive transport with the parameters $\nu_0 = 5 \times 10^{5}$, $\alpha = 0.5$, $\nu_c = \nu_l^f = \nu_l^t = 0$ and $T_C = T_D = 0$ and a BLL initial condition with $\alpha_{BLL} = 10$, (all non-dimensionalised).}
\label{fig:number_TOF_COMP}
\end{figure}

\begin{figure} 
\centering
	\includegraphics[width = \columnwidth]{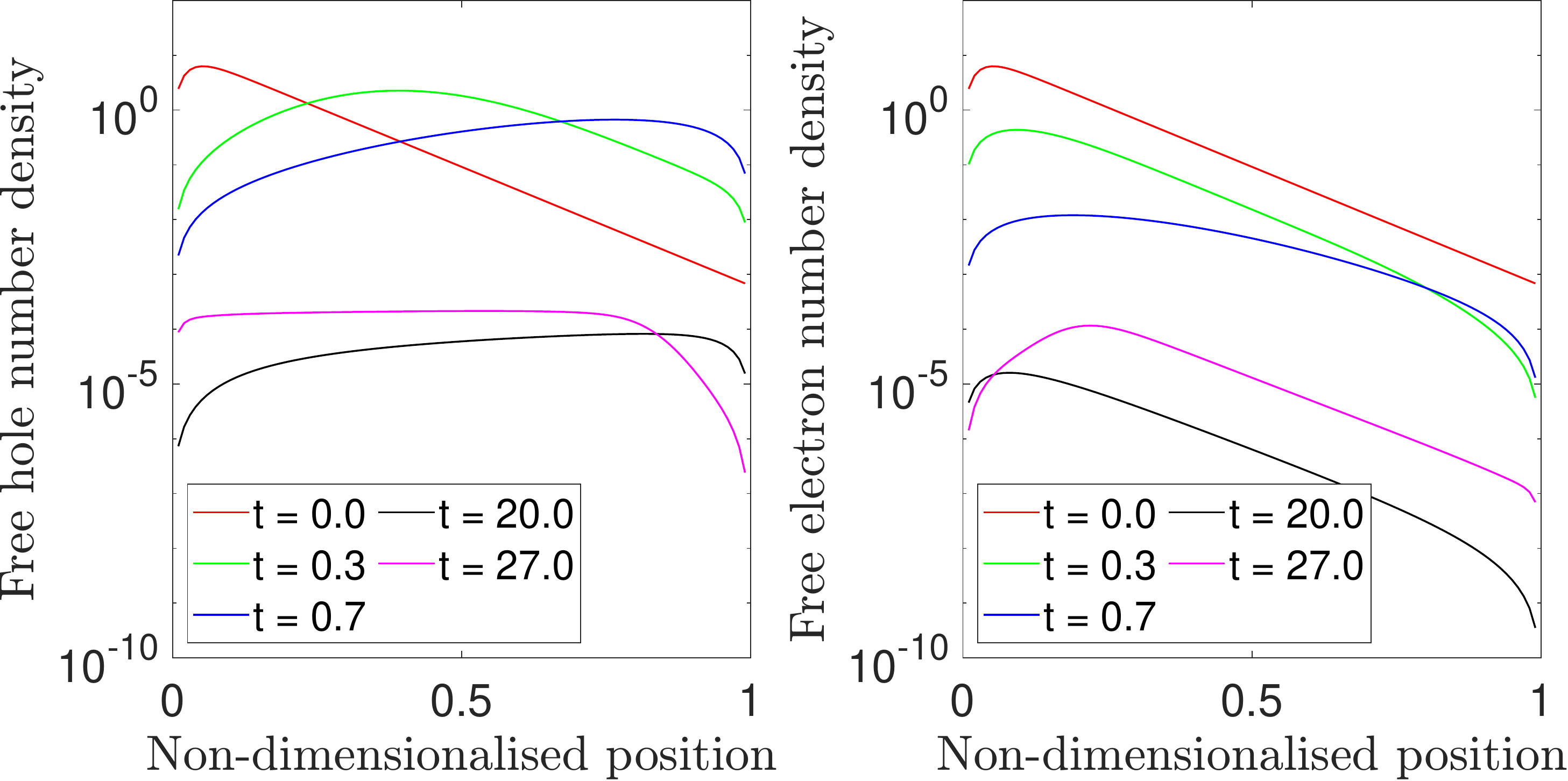}
\caption{Free particle number density when an AC field with $\omega = 0.02\pi$, is applied to a semiconductor undergoing dispersive transport with the parameters $\nu_0 = 5 \times 10^{5}$, $\alpha = 0.5$, $\nu_c = \nu_l^f = \nu_l^t = 0$ and $T_C = T_D = 0$ and a BLL initial condition with $\alpha_{BLL} = 10$, (all non-dimensionalised).}
\label{fig:number_TOF_NUMBD}
\end{figure}

The results of these simulations are presented in \figref{fig:number_TOF_COMP} and \figref{fig:number_TOF_NUMBD}. We see that the behaviour of the individual charge carriers is quite distinct. In particular, due to the relatively large initial number density close to the boundary and the opposite direction of acceleration there is approximately one order of magnitude difference in the total number of particles after one transit time. At later times we do still observe an increase in the number of particles at approximately the same time, reflecting the dependence of this property on the diffusion coefficient. 

It is also instructive to consider the number density evolution over time. We observe that the holes advect across the device within one transit time. On the other hand, there is less change in the shape of the number density of the electrons initially as they are advecting to the ``closest'' available boundary. The number density at later times displays the detrapping of particles (combined with some advection). The holes are detrapped more evenly (due to their advection across the device, smoothing the result out) whilst electron detrapping more closely reflects the initial condition.

We can alter the phase of the sinusoid and this will alter the results obtained. For instance a change in phase of $\pi$ would reverse the results for electrons and holes. This may prove very useful in designing an experiment. If we consider a material where the trapping rates of holes and electrons differ and if it were possible to perform multiple trials wherein the phase of the field was continuously changed it would be possible to observe the change in the magnitude of the conduction current with phase. This raises the possibility of constructing  the individual electron and hole profiles, although this is beyond the scope of this study.

The final point to note if we consider the application of our findings in an experimental setting is that the majority of the ``strange'' behaviour occurs at double the field frequency, aligning with the diffusion coefficient's variations, as seen in all the simulations presented above. This means that AC field experiments in the guise explored above will not reveal information about the trapping time. 

Overall, however, our results indicate that with certain materials an AC field experiment may well reveal qualitative information about trapping rates or be used to differentiate between electrons and holes. This could prove remarkably useful in the construction of more realistic models as well as device design. 

\section{Conclusions}
In this paper we have examined the generalised Boltzmann equation with collisions, trapping and detrapping defined using the BGK operator. We have determined the moments using standard differential equation techniques, extending the results of Stokes et al.~\cite{Stokes_Gen_Boltz}. The applications were then briefly discussed including the ratio of detrapping to trapping, the velocity moments and the temperature and diffusion tensors. In addition, the trapping operator was discussed and exponential detrapping was thoroughly explored and the results compared to those obtained by Robson and Makabe without trapping~\cite{Robson_Makabe_AC}. 

We proceeded to numerically explore the generalised diffusion equation and our results indicated several novel phenomena, specifically the increase in the number of particles when the diffusion coefficient is at its minima. The extraction of trapping rates for materials using AC fields is possible, however, current experimental set ups may prevent this at this stage. To develop an experiment using AC fields further study should be performed analysing the exact manner in which the information from AC fields can be used, specifically with different trapping rates for electrons and holes and with different field frequencies. It is felt that the Boltzmann equation with trapping and the generalised diffusion equation which arises provides a coherent and adequately general framework for analysing anomalous transport and will result in future breakthroughs from theoretical and experimental perspectives. 

\bibliography{Bibliography}

\end{document}